\documentclass[10pt,conference,compsocconf]{IEEEtran}
\usepackage{times}
\usepackage{graphicx}
\usepackage{caption}
\usepackage{txfonts}
\usepackage{array}
\ifCLASSINFOpdf
\else
\fi
\begin{document}
\pagenumbering{gobble}
%
\title{\textbf{\Large Formal verification of quantum communication protocols\\[-1.5ex] using
Petri nets}\\[0.2ex]}


\author{\IEEEauthorblockN{Sheng Zhang\IEEEauthorrefmark{1}, Yonghui Ma\IEEEauthorrefmark{2}, Chunning Meng\IEEEauthorrefmark{3} and
Haiping Wang\IEEEauthorrefmark{4} }
\IEEEauthorblockA{\IEEEauthorrefmark{1}\IEEEauthorrefmark{2}\IEEEauthorrefmark{3}\IEEEauthorrefmark{4}Department of Electronic Technology\\China Maritime Police Academy\\
Ningbo 315801 China}
\IEEEauthorblockA{\IEEEauthorrefmark{1} Email: huoxingren112@163.com}
\IEEEauthorblockA{\IEEEauthorrefmark{2} Email: 785820733@qq.com}
\IEEEauthorblockA{\IEEEauthorrefmark{3} Email: 2240529831@qq.com}
\IEEEauthorblockA{\IEEEauthorrefmark{4} Email: 123-whp@163.com}}

\maketitle

\begin{abstract}
This paper presents a new formal method for verification of quantum
communication protocols. By extending the symbolic system of Petri nets, we can define
quantum pure states in Petri-net settings. Therefore, it is possible to emerge a framework
from formalizing basic quantum phenomena, which are utilized to achieve communication
tasks. We also present an example of applying this framework to the modeling and analyzing
quantum communication protocols to show how it works.
\end{abstract}



%
\IEEEpeerreviewmaketitle

\section{Introduction}
Quantum communication, as one of the most promising technologies in future, is well known for it providing a way to achieve many tasks that are impossible purely by classical means, such as quantum dense coding\cite{bennett1992communication,mattle1996dense} and teleportation\cite{bennett1993teleporting,boschi1998experimental}. Among various quantum communication protocols, quantum key distribution\cite{Bennett1984Quantum} has become the focus of research all over the world  and achieves a huge success. However, security proof of such protocols\cite{Mayers1998Mayers,Shor2000Simple} remains to be the biggest problem, which prevents them from serving the up-to-date information systems in an acceptable cost. Therefore, formal methods were introduced to model, analyze and verify quantum protocols as they are applied to the classical cryptographic protocols\cite{Nagarajan2005An,M2006Relations,Feng2007Probabilistic,Feng2011Bisimulation,Feng2012Symbolic,Ying2009An,Deng2012Open,Ad2007A}. In 2005, Nagarajan et al. opened the door of adopting process calculi\cite{Milner1999Communicating} to quantum cryptographic protocols\cite{Nagarajan2005An}. Later, qCCS, where a quantum protocol is formalized as configuration $<P, \rho>$, have been proposed\cite{M2006Relations,Feng2007Probabilistic,Feng2011Bisimulation,Feng2012Symbolic,Ying2009An,Deng2012Open,Ad2007A}. Recently, Kubota et al. claimed a way to achieve semi-automated verification of security proofs of quantum cryptographic protocols in the qCCS framework\cite{Kubota2016Semi}. Independently, PRISM model checker\cite{Kwiatkowska2011PRISM} was employed to analyze the security of BB84\cite{Kim2004First}, however, the weakness is also intuitive, it only takes into account the simplest attacking strategy, i.e., intercept and resend attack. Recently, it was reported that more attack strategies are allowed in the PRISM settings\cite{Yang2016The}.

Although formal verifications of quantum cryptographic protocols have attracted most of the attentions, other quantum communication protocols were also investigated using formal methods. For example, Feng et al. formalized quantum teleportation and super dense coding protocols in qCCS\cite{Feng2011Bisimulation}. Here, we will continue the work by introducing a new formal method, Petri nets\cite{murata1989petri}, to modeling quantum communication protocols. Petri nets are a graphical and mathematical modeling tool applicable to many systems. It is often used to describe and analyze information processing systems that are characterized as being concurrent, asynchronous, distributed, parallel, nondeterministic, and stochastic. Obviously, quantum communication protocols are of such systems.

In this paper, we extend the symbolic system of Petri nets by a definition, which connects a quantum state with the elements of a typical Petri net. 
Next, we formalize several significant quantum phenomena, such as quantum interference, quantum entanglement, quantum measurement, and quantum Zeno
effect, in order to construct a new framework. These models act as the fundamentals to describe various quantum communication protocols. As an example,
we present a Petri net model of Salih et al.'s protocol, referred to as SLAZ2013\cite{salih2013protocol}. The main contribution of our work is to bridge
the gap between quantum protocols and Petri-net systems.

\section{Fundamentals}

Petri nets can be represented as follows:

  PN = (P, T, F, W, $M_{0}$ )

$P = \{ p_{1}, p_{2}, . . . , p_{m}\}$is a finite set of places,

$T = \{ t_{1}, t_{2}, . . . , t_{n}\}$ is a finite set of transitions,

$F \rightarrow (P \times T) U (T \times P)$ is a set of arcs (flow relation),

$W: f \rightarrow \{1, 2, 3, . . .\}$ is a weight function,

$M_{0}: P\rightarrow \{0, 1, 2, 3, . . .\}$ is the initial marking,

$P\bigcap T = \emptyset$  and  $P\bigcup T = \emptyset$ .

In order to describe quantum protocols in Petri-nets settings, we need to map the quantum states to Petri-nets symbolic systems.

\newtheorem{theorem}{Definition}
\begin{theorem}[Quantum state]
A quantum state, $|\Psi \rangle=\sum^{n}_{i=1}|\phi_{i}\rangle$, in Petri-net settings is described as a finite set of places ,$\tilde{p}=\{p_{1},p_{2},\cdots,p_{n}\}$, by the following mapping,
\begin{equation}\label{eq1}
Q: |\Phi(x)\rangle\mapsto \tilde{p},
\end{equation}
if we set $|C_{i}|^{2}=kM_{0}^{2}(p_{i})$, where $k$ is a positive number.

We should point out that, in real applications, the value of $k$ is most probably related to the computation complexity. For an instance, if $C_{i}$ is a real number, for $k=1$, the probability amplitude of the eigenstate $|\phi_{i}\rangle$ is equal to the token number of the place $p_{i}$. This definition directly bridges the quantum systems and Petri nets. Now, it is easy to formulate quantum phenomena in the Petri-nets framework.

\end{theorem}

\newtheorem{corollaries}{Corollaries}
\begin{corollaries}[Quantum inteference]
It is a natural application of the definition, since the function $M(p)$ is linear. Therefore, quantum interference can be expressed by

\begin{equation}\label{eq2}
|\Psi \rangle=|\phi_{1}\rangle+|\phi_{2}\rangle  \Longleftrightarrow M(p)=M_{1}(p)+M_{2}(p).
\end{equation}
\end{corollaries}

\begin{corollaries}[Quantum measurement]
Defining an operator $\hat{M}_{B}$ for a quantum state denoted by $<Q: |\Phi(x)\rangle\mapsto \tilde{p}>$,  we have $\hat{M}_{B}\tilde{p}=p_{i}$, such that $Prob\{p=p_{i}\}=kM^{2}(p_{i})$, if the eigenstates of $\tilde{p}$ are chosen as the measurement basis. Our solution to quantum measurement directly represents the essence of quantum collapse induced by disturbed measurement, and it can be easily applied to various applications, such as quantum key distribution. Fig. \ref{quant_measurement} is an example of simulating quantum measurement in a Petri net system.

\begin{figure}[!t]
\centering
\includegraphics[width=2.75in,height=2.75in]{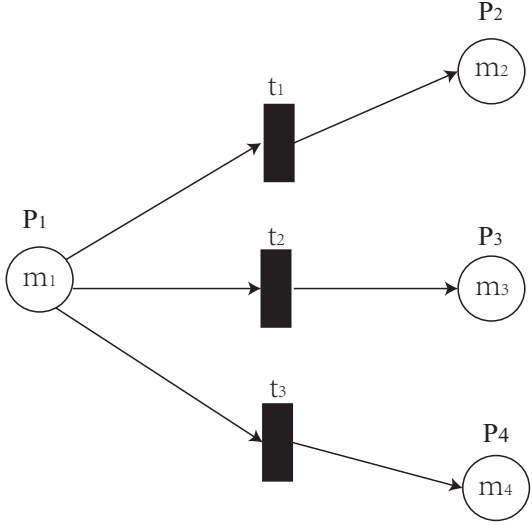}
\caption{A Petri net representing quantum measurement. The weights of arcs $F\rightarrow\{p_{1}\times t_{1}, p_{1}\times t_{2}, p_{1}\times t_{3}, t_{1}\times p_{2}, t_{2}\times p_{3}, t_{3}\times p_{4}\}$ are set as $W: f \rightarrow \{1, 1, 1, 1/\sqrt{3}, 1/\sqrt{3}, 1/\sqrt{3}\}$. $M_{0}: P\rightarrow \{1, 0, 0, 0\}$ is the initial marking.}\label{quant_measurement}
\end{figure}

\end{corollaries}

\begin{corollaries}[Quantum entanglement]
 Quantum entanglement is a striking phenomenon explored to achieve many tasks in quantum information. Here, we can use a Petri net to draw the picture of quantum entanglement as showed in Fig. \ref{quant_entanglement}. Explicitly, we can use a Petri net to describe a two-qubit entangled system, say, a triplet state $|\psi^{+}>=1/\sqrt{2}(|10>+|01>)$.
\begin{figure}[!t]
  \centering
  \includegraphics[width=3.05in,height=3.75in]{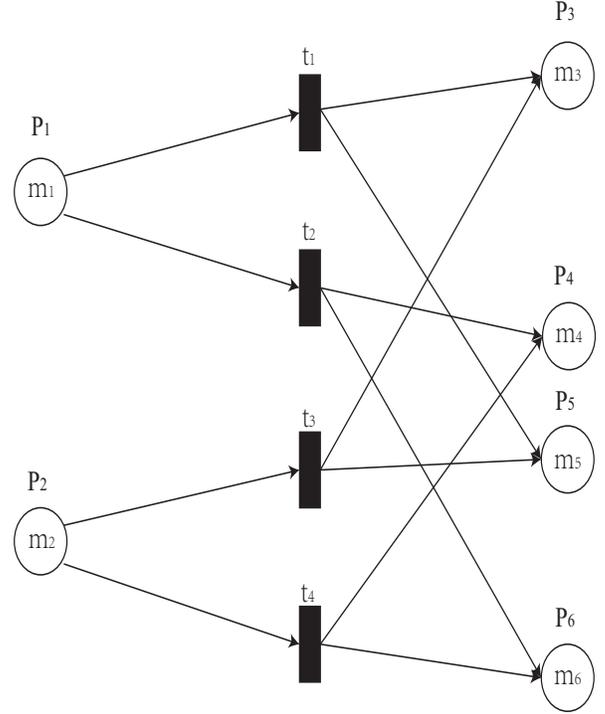}\\
  \caption{A Petri net representing quantum entanglement. The weights of the arcs are set as 1, and $M_{0}: P\rightarrow \{1, 1, 0, 0, 0, 0\}$ is the initial marking.}\label{quant_entanglement}
\end{figure}

$PN_{|\psi^{+}>}=(P, T, F, W, M_{0})$ corresponds to a triplet state, such that

$\cdot P1=\cdot P2=\varnothing$,

$P3\cdot=P4\cdot=P5\cdot=P6\cdot=\varnothing$,

$P1\cdot=\{t1,t2\},P2\cdot=\{t3,t4\}$,

$\cdot P3=\cdot P5=\{t1,t3\},\cdot P4=\cdot P6=\{t2,t4\}$,

$\cdot t1=\cdot t2=\{P1\},\cdot t3=\cdot t4=\{P2\}$,

$t1\cdot=t3\cdot=\{P3,P5\},t2\cdot=t4\cdot=\{P4,P6\}$.

Now, let us see how quantum entanglement is drawn in this picture:

In the beginning, the system should be initialized before it is triggered, i.e., $m_{1}=M_{0}(p_{1})=1, M_{0}(p_{2})=M_{0}(p_{3})=M_{0}(p_{4})=M_{0}(p_{5})=M_{0}(p_{6})=0$. Consequently, according to the rules of Petri nets systems, two possible results are going to be observed: (I) If the token flows to the place $p_{4}$, i.e., $m_{1}=0$, $M_{0}(p_{1})[t2\rangle$, $m_{4}=M(p_{4})=M_{0}(p_{4})+1=1$, one immediately observes $m_{6}=M(p_{6})=1$; (II)If the token flows to the place $p_{3}$, i.e., $m_{1}=0$, $M_{0}(p_{1})[t1\rangle$, $m_{3}=M(p_{3})=M_{0}(p_{3})+1=1$, one immediately observes $m_{5}=M(p_{5})=1$. Interestingly, cases (I) and (II) can be translated to the components $|10>$ and $|01>$ of $|\psi^{+}>$.

\end{corollaries}

\begin{corollaries}[Quantum Zeno effect]
Quantum Zeno effect\cite{Misra1977The} is a phenomenon in quantum physics where observing a particle prevents it from decaying as it would in the absence of the observation. A good example of employing quantum Zeno effect to quantum communication protocols is presented in Ref.\cite{salih2013protocol}, where a $N$-cycle interferometer is introduced. Quantum Zeno effect is seen in the following scenario: After n cycles, the evolution of the initial state $|10>$ is expressed by
\begin{equation}\label{eq3}
 |10> \rightarrow \cos^{n-1} \theta(\cos \theta |10>+\sin \theta |01>).
\end{equation}
In the end, $|10>$ will be detected with a probability of $\cos^{2N}\approx 1$. Fig.\ref{Zeno} is a Petri-net system correlated to this example, place $p_{9}$ stores the total cycle number denoted by $N$, $p_{11}$ and $p_{12}$ represent the final states $|10>$ and $|01>$, respectively. In other words, two quantum states, $<Q: |10>\rightarrow p_{11}>$ and $<Q: |01>\rightarrow p_{12}>$, can be found in this Petri net. Running this system, we obtain $kM^{2}(p_{11})\approx 1$, which is consistent with the result reduced from Eq. (\ref{eq3}).

\begin{figure}
  \centering\includegraphics[width=3.75in,height=3.75in]{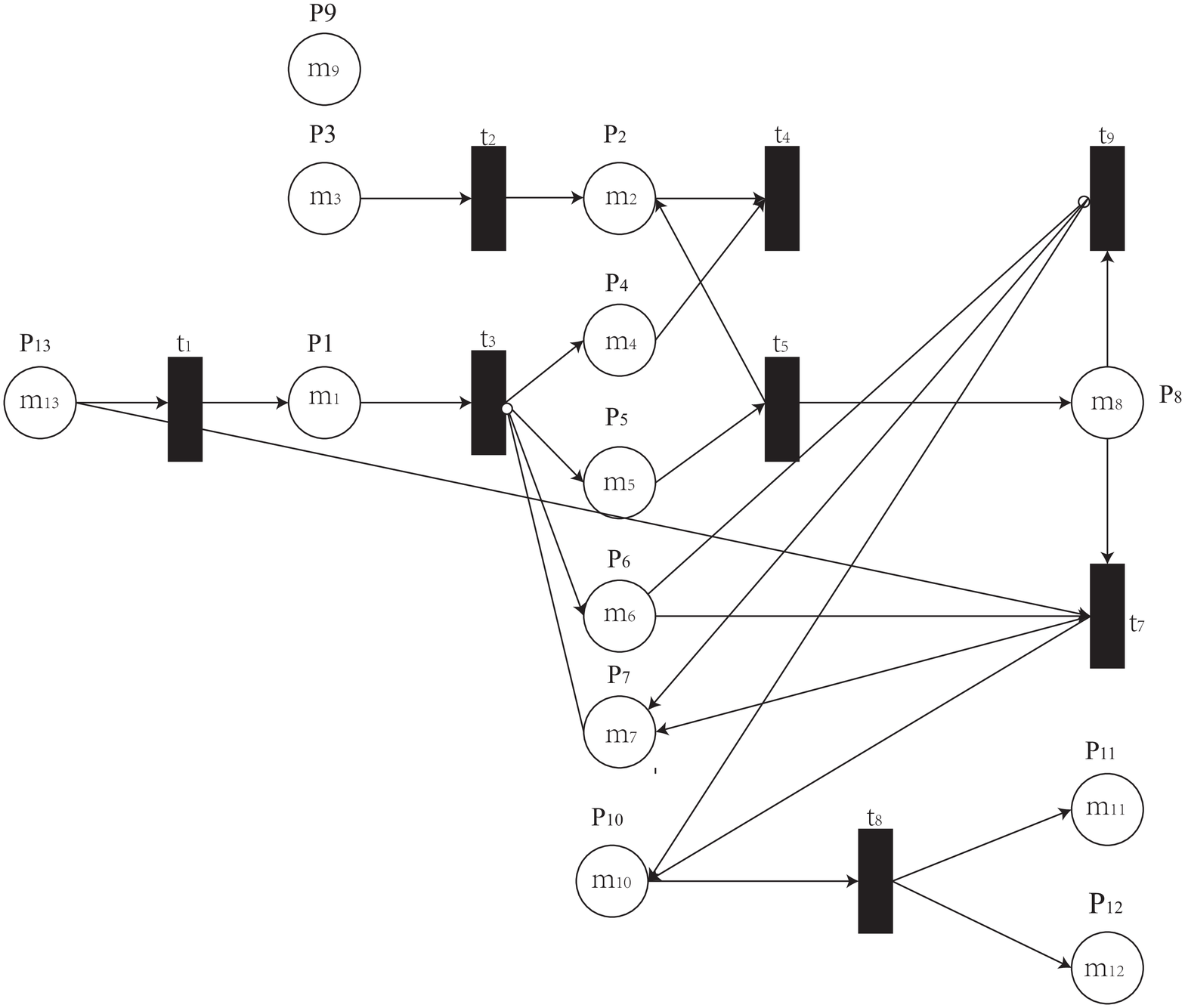}\\
  \caption{A Petri net representing quantum Zeno effect. The weights of arcs are set to be 1 except $\{p_{7}\times t_{3}, t_{2}\times p_{2}, p_{2}\times t_{4}, t_{5}\times p_{2}, t_{3}\times p_{6}, p_{10}\times t_{6}, t_{6}\times p_{11}, t_{6}\times p_{12}\}$ being $\{m_{9}-2, \cos(\pi/2m_{9})\times 10^{14}, m_{2}, m_{6}, \cos(\pi/2m_{9})\ast m_{2}, m_{9}-2, \cos(\pi/2m_{9})\ast m_{2}, \sin(\pi/2m_{9})\ast m_{2}\}$. Here, we set $k=10^{-28}$ to coordinate the probability amplitude with the token number, i.e., $C_{i}=10^{-14}M(p_{i})$. $M_{0}: P\rightarrow \{0, 0, 1, 0, 0, 0, 0, 0, N, 0, 0, 0, 1\}$ is the initial marking ($N$ is a prefixed natural number indicating the cycle number). }\label{Zeno}
\end{figure}
\end{corollaries}

\label{fd}

\section{Verifying the direct counterfactual quantum communication protocol by Petri nets}

In this section, we present a model of a typical quantum communication protocol proposed by Salih et al.\cite{salih2013protocol}, in order to verify the counterfactuality rate of this protocol. The model is comprised of two Petri nets, which represent two work modes related to Bob passing and blocking the photon, respectively. It is partially adapted from the prototype illustrated in Fig. \ref{Zeno}, since quantum Zeno effect is introduced to this protocol and takes place in both modes. Without specially clarification, parameter $k$ is set to be $10^{-28}$ as it was in the previous section.

\subsection{Model}
When the protocol works in the blocking mode, i.e., logic "1" is transmitted, quantum Zeno effect is observed in the inner cycle. Due to the quantum interference in the outer cycle, the photon will trigger the corresponding detector (denoted here by $D_{2}$). It implies that the initial state $|10>$ finally evolves to $|01>$.

\begin{figure}
  \centering\includegraphics[width=3.75in,height=3.75in]{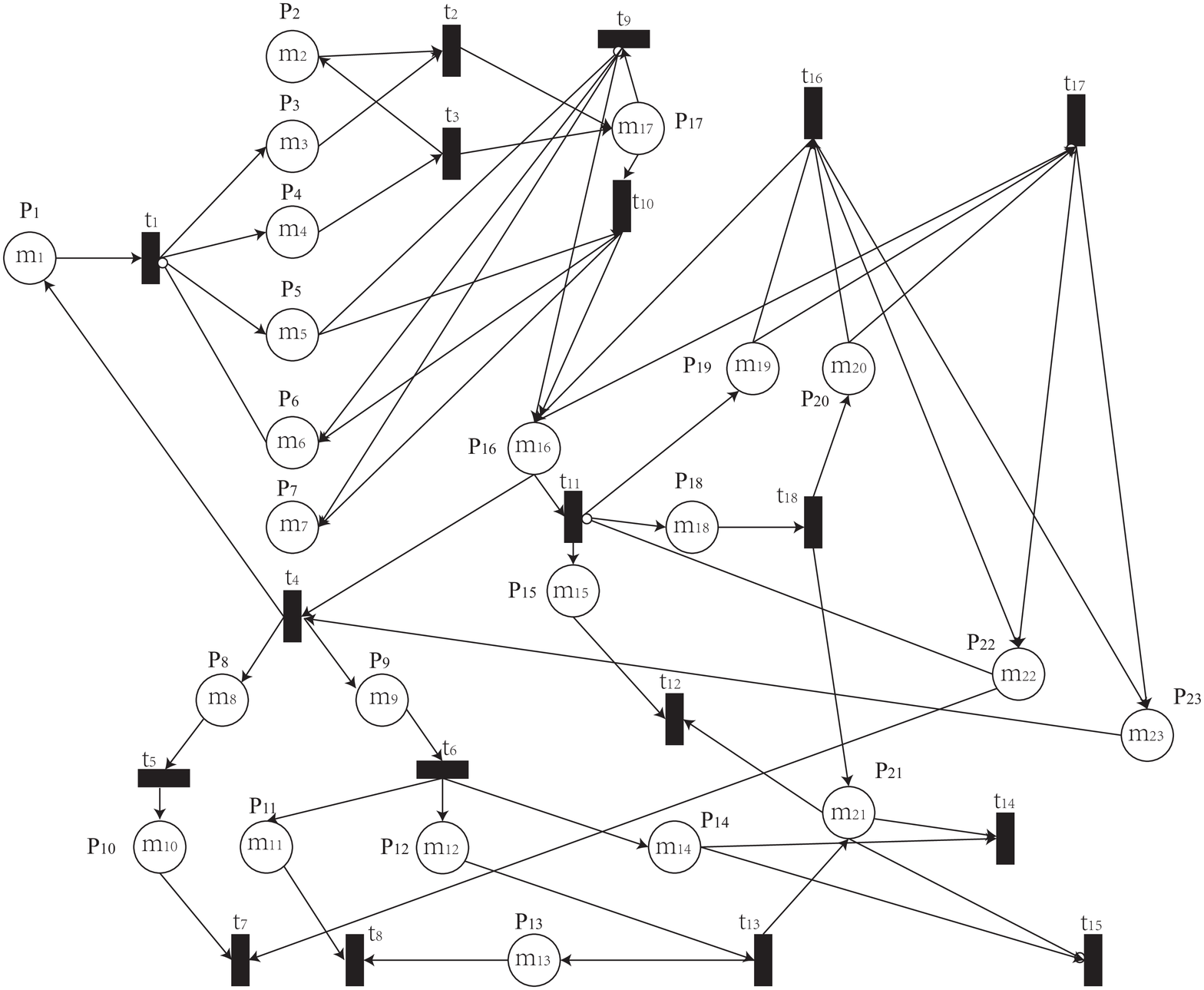}\\
  \caption{A Petri net modeling the blocking mode.}\label{c0}
\end{figure}

A Petri net model of this mode is shown in Fig. \ref{c0}. In order to show the principles, we divide the net into three subsystems denoted by $PN^{0}_{1}$, $PN^{0}_{2}$ and $PN^{0}_{3}$. $PN^{0}_{1}$ carries out the function of the inner cycle, and $PN^{0}_{2}$ and $PN^{0}_{3}$ respond to the outer cycle.

$PN^{0}_{1}$ is described below.

Places: $P= \{ p_{15}, p_{16}, p_{18}, p_{19}, p_{20}, p_{21}, p_{22}, p_{23}\}$;

Transitions: $T = \{ t_{11}, t_{12}, t_{16}, t_{17}, t_{18}\}$;

The Weights of the arcs are set to 1 except the followings:

$N$ (a prefixed natural number indicating the inner cycle number) for $P_{22}\times t_{11}$ and $P_{23}\times t_{4}$;

$\cos(\pi/2N)\ast m_{21}$ for $t_{11}\times P_{19}$;

$m_{19}$ for $P_{19}\times t_{16}$;

$m_{19}$ for $t_{18}\times P_{21}$;

$m_{21}$ for $P_{21}\times t_{12}$.

The resulted state after $N$ cycles is expressed by $<Q: |010>\rightarrow p_{21}>$, which is input to the next outer cycle referred to $PN^{0}_{2}$ and $PN^{0}_{3}$.

$PN^{0}_{2}$ is described below.

Places: $P= \{ p_{1}, p_{2}, p_{3}, p_{4}, p_{5}, p_{6}, p_{7}, p_{17}\}$;

Transitions: $T = \{ t_{1}, t_{2}, t_{3}, t_{9}, t_{10}\}$;

The Weights of the arcs are set to 1 except the followings:

$M$ (a prefixed natural number indicating the outer cycle number) for $P_{6}\times t_{1}$;

$\cos(\pi/2M)\ast m_{2}-\sin(\pi/2M)\ast m_{21}$ for $t_{1}\times P_{5}$;

$m_{2}$ for $P_{2}\times t_{2}$;

$m_{5}$ for $t_{3}\times P_{2}$.

The photon appears in the left hand side arm of the interferometer after $m$ cycles is expressed by $<Q: |100>\rightarrow p_{2}>$.

$PN^{0}_{3}$ is described below.

Places: $P= \{ p_{8}, p_{9}, p_{10}, p_{11}, p_{12}, p_{13}, p_{14}, p_{21}, p_{22}\}$;

Transitions: $T = \{ t_{4}, t_{5}, t_{6}, t_{7}, t_{8}, t_{13}, t_{14}, t_{15}\}$;

The Weights of the arcs are set to 1 except the followings:

$\sin(\pi/2M)\ast m_{2}-\cos(\pi/2M)\ast m_{21}$ for $t_{6}\times P_{11}$;

$m_{28}$ for $P_{11}\times t_{8}$;

$m_{11}$ for $t_{13}\times P_{21}$;

$m_{21}$ for $P_{21}\times t_{14}$;

$N$ for $P_{22}\times t_{7}$.

The photon appears in the right hand side arm after $m$ cycles is expressed also by $<Q: |010>\rightarrow p_{21}>$. Consequently, the photon triggers detector $D_{2}$ with a probability of $kM^{2}(p_{21})$, leading to a result state $|010>$. Note that, in the blocking mode, quantum Zeno effect occurs in the inner cycle.

When the protocol works in the passing mode, i.e., logic "0" is transmitted, quantum Zeno effect is observed in the outer cycle. As a result, the photon will trigger the other detector (denoted by $D_{1}$). It implies that the initial state $|10>$ remains unchanged after being continuously observed.

\begin{figure}
  \centering\includegraphics[width=3.75in,height=3.75in]{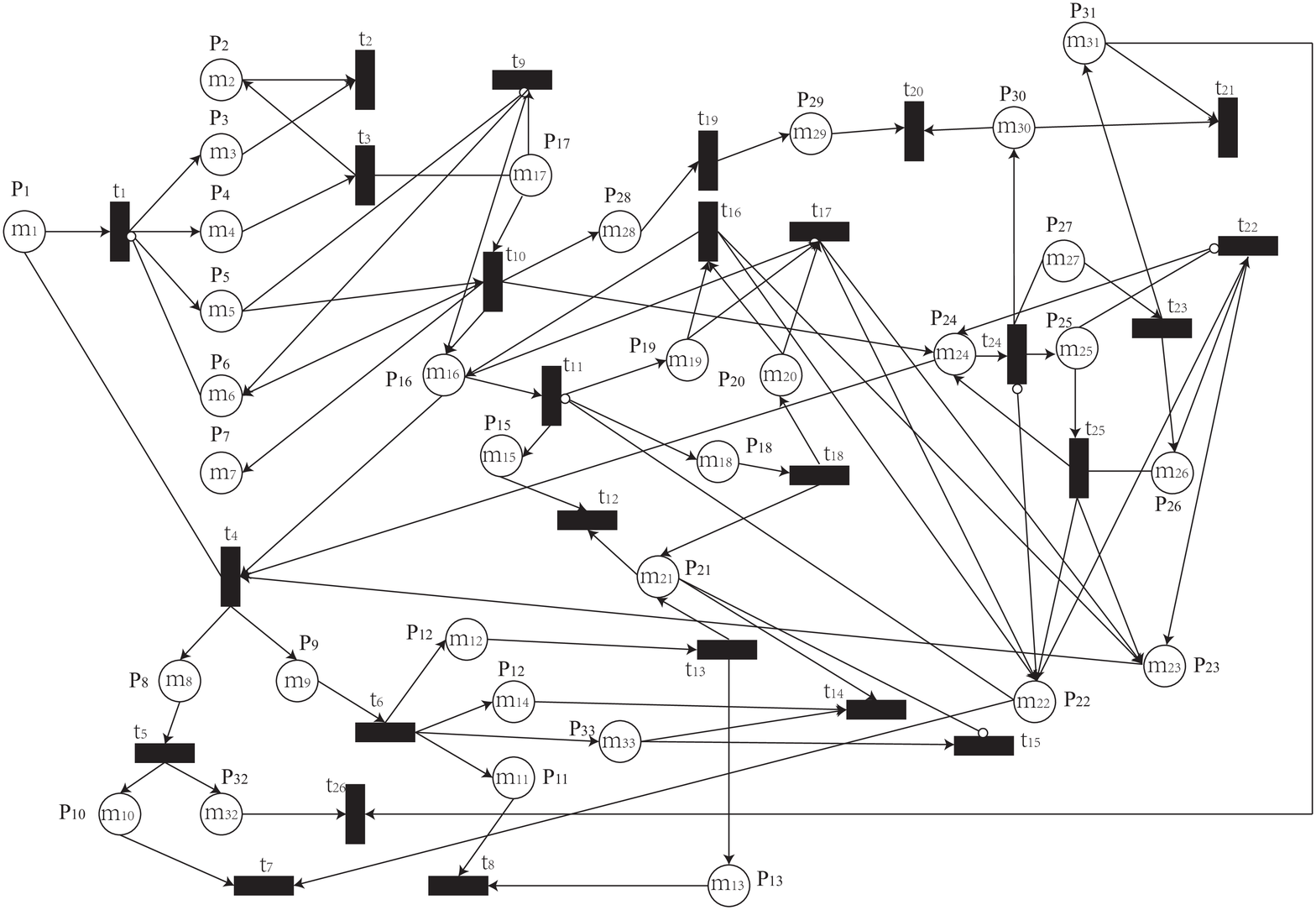}\\
  \caption{A Petri net modeling the passing mode.}\label{c1}
\end{figure}

A Petri net model of this mode is shown in Fig. \ref{c1}. Similarly, it is divided into four subsystems denoted by $PN^{1}_{1}$, $PN^{1}_{2}$, $PN^{1}_{3}$ and $PN^{1}_{4}$. $PN^{1}_{1}$ and $PN^{1}_{2}$ together carry out the function of the inner cycle with $PN^{1}_{3}$ and $PN^{1}_{4}$ responding to the outer cycle. According to principle of the protocol, Fig. \ref{c1} can be easily adapted from Fig. \ref{c0}.

$PN^{1}_{1}$ is described below.

Places: $P= \{ p_{15}, p_{16}, p_{18}, p_{19}, p_{20}, p_{21}, p_{22}, p_{23}\}$;

Transitions: $T = \{ t_{11}, t_{12}, t_{16}, t_{17}, t_{18}\}$;

The Weights of the arcs are set to 1 except the followings:

$2N$ for $P_{22}\times t_{11}$ and $P_{23}\times t_{4}$;

$\cos(\pi/2N)\ast m_{21}-\sin(\pi/2N)\ast m_{31}$ for $t_{11}\times P_{19}$;

$m_{19}$ for $P_{19}\times t_{16}$;

$m_{19}$ for $t_{18}\times P_{21}$;

$m_{21}$ for $P_{21}\times t_{12}$.

The resulted state after $N$ cycles is expressed by $<Q: |010>\rightarrow p_{21}>$, which is input to the next outer cycle, i.e., $PN^{1}_{3}$ and $PN^{1}_{4}$.

$PN^{1}_{2}$ is described below.

Places: $P= \{ p_{22}, p_{23}, p_{24}, p_{25}, p_{26}, p_{27}, p_{28}, p_{29}, p_{30}, p_{31}\}$;

Transitions: $T = \{ t_{10}, t_{19}, t_{20}, t_{21}, t_{22}, t_{23}, t_{24}, t_{25}, t_{26}\}$;

The Weights of the arcs are set to 1 except the followings:

$\sin(\pi/2N)\ast m_{2}+\cos(\pi/2N)\ast m_{31}$ for $t_{24}\times P_{25}$;

$m_{31}$ for $P_{31}\times t_{21}$;

$m_{25}$ for $t_{23}\times P_{31}$;

$m_{25}$ for $P_{25}\times t_{25}$.

The resulted state after $N$ cycles is expressed by $<Q: |001>\rightarrow p_{31}>$, which is input to the next outer cycle, i.e., $PN^{1}_{3}$ and $PN^{1}_{4}$.

$PN^{1}_{3}$ is described below.

Places: $P= \{ p_{1}, p_{2}, p_{3}, p_{4}, p_{5}, p_{6}, p_{7}, p_{17}, p_{28}\}$;

Transitions: $T = \{ t_{1}, t_{2}, t_{3}, t_{9}, t_{10}\}$;

The Weights of the arcs are set to 1 except the followings:

$M$ for $P_{6}\times t_{1}$;\textbf{...}

$\cos(\pi/2M)\ast m_{2}-\sin(\pi/2M)\ast m_{21}$ for $t_{1}\times P_{5}$;

$m_{2}$ for $P_{2}\times t_{2}$;

$m_{5}$ for $t_{3}\times P_{2}$.

The photon appears in the left hand side arm after $m$ cycles is expressed by $<Q: |100>\rightarrow p_{2}>$.

$PN^{1}_{4}$ is described below.

Places: $P= \{ p_{8}, p_{9}, p_{10}, p_{11}, p_{12}, p_{13}, p_{14}, p_{21}, p_{31}, p_{32}, p_{33}\}$;

Transitions: $T = \{ t_{4}, t_{5}, t_{6}, t_{7}, t_{8}, t_{13}, t_{14}, t_{15}, t_{26}\}$;

The Weights of the arcs are set to 1 except the followings:

$\sin(\pi/2M)\ast m_{2}-\cos(\pi/2M)\ast m_{21}$ for $t_{6}\times P_{11}$;

$m_{28}$ for $P_{11}\times t_{8}$;

$m_{11}$ for $t_{13}\times P_{21}$;

$m_{21}$ for $P_{21}\times t_{14}$;

$m_{31}$ for $P_{21}\times t_{36}$;

$N$ for $P_{22}\times t_{7}$.

The photon appears in the right hand side arm after $m$ cycles is expressed also by $<Q: |010>\rightarrow p_{21}>$. Since quantum Zeno effect takes place in the outer cycle, the photon triggers detector $D_{1}$ with a probability of $kM^{2}(p_{2})$, leading to a result state $|100>$.

\subsection{Verification results}

In order to verify the correctness, the model is run in the Petri nets environment with different values of $N$ and $M$, thus a list of results of the counterfaulity rates, which equal to the detection probabilities as well, is obtained.  Meanwhile, the same terms are computed out in MATLAB. Comparing the data , we find that the computational results produced by the Petri net model are completely consistent with those from MATLAB. We have listed below a part of the results in Tab. \ref{table_1} and \ref{table_2}.

\begin{table}[!t]
\renewcommand{\arraystretch}{1.3}
\caption{The counterfactuality rate for the passing mode}
\label{table_1}
\begin{tabular}{p{1cm}<{\centering}p{1cm}<{\centering}p{1cm}<{\centering}p{1cm}<{\centering}p{1cm}<{\centering}p{1cm}<{\centering}}
\hline
&$M=25$&$M=50$&$M=75$&$M=100$&$M=150$\\
\hline
N=320&0.906 &0.952&0.968 &0.976 &0.984 \\
N=500&0.906 &0.952 &0.968 &0.976 &0.984 \\
N=1250&0.906 &0.952 &0.968 &0.976 &0.984 \\
N=2500&0.906 &0.952 &0.968 &0.976 &0.984 \\

\end{tabular}
\end{table}

\begin{table}[!t]
\renewcommand{\arraystretch}{1.3}
\caption{The counterfactuality rate for the blocking mode}
\label{table_2}
\centering
\begin{tabular}{p{1cm}<{\centering}p{1cm}<{\centering}p{1cm}<{\centering}p{1cm}<{\centering}p{1cm}<{\centering}p{1cm}<{\centering}}
\hline
&$M=25$&$M=50$&$M=75$&$M=100$&$M=150$\\
\hline
N=320&0.912  &0.831  &0.758 &0.693 &0.582\\
N=500&0.943 &0.887 &0.836 &0.788 &0.702 \\
N=1250&0.977 &0.953 &0.930 &0.908 &0.865 \\
N=2500&0.997 &0.994 &0.991 &0.988 &0.982 \\

\end{tabular}
\end{table}

\section{Discussion}
We should point out that the presented framework has limitations to describe more general quantum phenomena. For example, the model of quantum entanglement presented in Sec.\ref{fd} is only suitable for a Bell state. Fortunately, it is not difficult to generalize it to a $N$-particle entangled system. Thus, it is necessary to fertilize the framework by formalizing more quantum mechanics serving the quantum information in our future work.

We have presented a good example of modeling quantum communication protocols in Petri-net settings, it is still open that whether it is available to model and verify quantum cryptographic protocols using Petri-net models, since the security aspect of these protocols is more needed to be investigated. However, it is a big challenge to construct an automated reasoning system that performs the security proof in Petri nets. Also, we need to formalize various eavesdropping strategies, such as collective attacks, as is left for our future work. Another challenge is how to verify composed systems, where quantum communication protocols and classical ones are combined with each other to provide network services. Indeed, verifications of the composable security is also expected.
\section{Conclusion}

In this paper, we presented a Petri-net framework for modeling and analyzing quantum communication protocols, which opens a new door of verifying quantum protocols by formal methods. In order to make it possible to describe quantum protocols, we extended the Petri-net symbolic system using a mapping from Hilbert space to the place set. Therefore, quantum states can be defined in the Petri-net settings. Typical quantum phenomena were also formalized, so that they provide us fundamental tools to model quantum protocols, where these phenomena are usually used to achieve amazing tasks that are definitely impossible by classical means. At last, we presented a Petri-net model of SLAZ2013 protocol to show how this framework works.

\addcontentsline{toc}{chapter}{Acknowledgment}
\section*{Acknowledgment}
This work was supported by the National Natural Science Fundation of China with the project number 61300203.

\captionsetup{font={footnotesize,rm},justification=centering,labelsep=period}%

\bibliographystyle{IEEEtran}
\bibliography{bibtemplate_samples}

\end{document}